# Normal and Anomalous Fluctuation Relations for Gaussian Stochastic Dynamics

## A V Chechkin[1,3], F Lenz[2] and R Klages[2]

[1] Institute for Theoretical Physics NSC KIPT, Akademicheskaya st. 1, Kharkov 61108, Ukraine
[2] Queen Mary University of London, School of Mathematical Sciences, Mile End Road, London E1 4NS, UK

E-mail: achechkin@kipt.kharkov.ua, f.lenz@qmul.ac.uk, and r.klages@qmul.ac.uk

**Abstract.** We study transient work Fluctuation Relations (FRs) for Gaussian stochastic systems generating anomalous diffusion. For this purpose we use a Langevin approach by employing two different types of additive noise: (i) internal noise where the Fluctuation-Dissipation Relation of the second kind (FDR II) holds, and (ii) external noise without FDR II. For internal noise we demonstrate that the existence of FDR II implies the existence of the Fluctuation-Dissipation Relation of the first kind (FDR I), which in turn leads to conventional (normal) forms of transient work FRs. For systems driven by external noise we obtain violations of normal FRs, which we call anomalous FRs. We derive them in the long-time limit and demonstrate the existence of logarithmic factors in FRs for intermediate times. We also outline possible experimental verifications.

**Keywords**: stochastic processes (theory), fluctuations (theory), diffusion, stochastic particle dynamics (theory).

## Contents

1. Introduction
2. Generalized Langevin equation and transient work fluctuation relation for stochastic Gaussian systems with constant force
3. Internal noise case: normal fluctuation relations
4. External noise case: MSD and fluctuation relations
   4.1. Persistent external noise
   4.2. Antipersistent external noise
5. Logarithmic corrections at intermediate times
   5.1. Persistent external noise
   5.2. Antipersistent external noise
6. Conclusions

## 1. Introduction

*Fluctuation Relations* (FRs) belong to the rare laws of statistical physics that are valid very far from equilibrium, as has been confirmed for a great variety of systems; see [1] and further references therein. They can be understood as large-deviation symmetry properties of the probability density functions (PDFs) of statistical physical observables in nonequilibrium situations. One fundamental form of them, often referred to as *Fluctuation Theorems*, emerged as nonlinear Fluctuation-Dissipation Relations within a Hamiltonian statistical mechanical framework [2] and by generalizing the Second Law of Thermodynamics to thermostated chaotic dynamical systems in nonequilibrium [3]–[5]. This form was also found to hold for general Markov processes [6], [7]. Another basic type extends an equilibrium relation between work and free energy to nonequilibrium [8]. These fundamental forms can partially be derived from other, more general FRs as special cases [9]–[12]. Many of these relations have been verified in experiments on small systems [13]–[17].

---

[3] Institute of Physics and Astronomy, University of Potsdam, 14476 Potsdam-Golm, Germany

In this letter we focus on *transient work* FRs, which define an important, generic type of FRs within the first form [18]. They characterize the PDF $p(W,t)$ of the production of work $W$ over time $t$ for a system coupled to a thermal reservoir at temperature $T$ and driven by an external force, which starts in a given initial state (either defined by an initial condition for a single particle, or by an initial distribution for an ensemble) by moving into a nonequilibrium steady state. In pioneering work by Evans et al. [4], [13] it has been shown analytically, by computer simulations, and experimentally that the following relation between positive and negative production of work holds (in case of time symmetric protocols for the external force):

$$\ln \frac{p(W,t)}{p(-W,t)} = \frac{W}{k_B T} \quad , \tag{1.1}$$

where $k_B$ is the Boltzmann constant. This form, which we call the *normal* FR, has then been confirmed for many other types of dynamics, and by many experiments [1].

An interesting generalization of normal FR, still keeping the functional form of Eq.(1.1), has been obtained for glassy dynamics by replacing the thermostat temperature $T$ through a suitably defined nonequilibrium *effective temperature* $T_{FR}$ [19], [20]. This generalization cross-links to findings of *anomalous dynamics*, where it was shown both analytically and numerically that the normal FR can be violated in various ways [21]–[27]. Any violation of Eq.(1.1) we call an *anomalous FR*. Anomalous dynamics refers to processes that do not obey the laws of "conventional" statistical physics [28]. Such dynamics is exemplified by diffusion processes where the long-time mean square displacement does not grow linearly in time, $\langle x^2(t) \rangle \propto t^\mu$ with $\mu = 1$ as for ordinary Brownian motion, but either subdiffusively with $\mu < 1$ or superdiffusively with $\mu > 1$. Anomalous transport phenomena have been detected both theoretically and experimentally in a wide variety of complex systems [29], [30], [28]. A non-universality of FRs for steady-state stochastic systems was also demonstrated in [31], [32] by studying dynamics under the influence of external asymmetric Poissonian shot noise.

In [24], [27] we tested work FRs for three generic types of anomalous dynamics: Lévy flights, long-correlated Gaussian processes, and time-fractional kinetics. Here we continue this line of research by checking work FRs for paradigmatic examples of stochastic Gaussian systems, corresponding to physical situations which yield four different types of anomalous diffusion. We model this dynamics by a Langevin stochastic differential equation with constant force and additive stationary noise obeying a Gaussian PDF. Two types of noise are considered: (i) internal noise where the *Fluctuation-Dissipation Relation of the second kind* (FDR II) is fulfilled, and (ii) external noise without FDR II. We find that for systems driven by internal Gaussian noise normal FRs always hold, whereas for systems driven by external Gaussian noise anomalous FRs are possible. To clarify the role of correlations in the noise, for each of the two types of noise we consider two typical noise autocorrelation functions (ACFs) with power law decay, exhibiting persistent and antipersistent behaviour, respectively.

## 2. Generalized Langevin equation and transient work fluctuation relation for stochastic Gaussian systems with constant force

Our starting point is the generalized Langevin equation for the overdamped random motion of a particle [34]

$$\int_0^t dt' \gamma(t-t') \frac{dx(t')}{dt'} = \frac{F}{m} + \xi(t) \quad , \tag{2.1}$$

where $F$ is a constant force, $m$ the mass of the particle and $\gamma(t)$ the friction kernel. The stationary random noise $\xi(t)$ obeys Gaussian statistics with zero mean, $\langle \xi(t) \rangle = 0$, where $\langle ... \rangle$ denotes the ensemble average over different realizations of the random force. Hence, the noise is completely characterized by its ACF $g(\tau) = \langle \xi(t) \xi(t') \rangle_{\tau = t-t'}$, which is an even function as specified below. Here we are interested in transient

FRs for the mechanical work $W = Fx$, which is identical to the heat for systems driven by a constant force (class A systems according to the classification of [33]). Therefore, the PDF $p(W,t)$ for the work is simply related to the PDF $f(x,t)$ of the position $x$ of the particle by $p(W,t) = F^{-1} f(W/F, t)$. Since the system defined by Eq.(2.1) is linear and Gaussian, the work PDF reads

$$p(W,t) = \frac{1}{\sqrt{2\pi\sigma_W^2(t)}} \exp\left\{-\frac{(W-\langle W \rangle)^2}{2\sigma_W^2(t)}\right\} \;, \tag{2.2}$$

where the mean $\langle W \rangle$ and the variance $\sigma_W^2(t)$ of work are related to the mean displacement (MD) $\langle x(t) \rangle$ and the mean squared displacement (MSD) $\sigma_x^2(t) = \langle (x(t) - \langle x(t) \rangle)^2 \rangle$ of the process $x(t)$ by

$$\langle W \rangle = F \langle x(t) \rangle \;, \qquad \sigma_W^2(t) = \langle (W(t) - \langle W(t) \rangle)^2 \rangle = F^2 \sigma_x^2(t) \;. \tag{2.3}$$

The transient FR for mechanical work is then given by

$$\ln \frac{p(W,t)}{p(-W,t)} = \frac{2\langle x(t) \rangle}{F\sigma_x^2(t)} W \;. \tag{2.4}$$

Thus, to check the work FR for the Gaussian system obeying Eq.(2.1) one only needs to calculate the MD and the MSD for the process $x(t)$. To do that, we now specify the properties of the noise ACF and consider two types of noise: internal and external. Since there is no Boltzmann equilibrium for systems with constant force, and in order to be consistent, we choose the simplest nonequilibrium initial condition $x(0) = 0$ in both cases.

## 3. Internal noise case: normal fluctuation relations

The ACF of internal noise is related to the friction kernel $\gamma(t)$ by the FDR II [34],

$$g(\tau) = \langle \xi(t)\xi(t') \rangle_{\tau = t-t'} = \frac{k_B T}{m} \gamma(t-t') \;. \tag{3.1}$$

In the special case of ordinary Brownian motion $\gamma(t-t') = 2\gamma\delta(t-t')$, $\gamma$ is the friction constant, and $\xi(t)$ is white noise, $\langle \xi(t)\xi(t') \rangle = 2D\delta(t-t')$, $D = k_B T \gamma / m$ (we adopt $\int_0^t \delta(\tau)d\tau = 1/2$). Here we consider the general case of non-Markovian noise.

Let us show that for systems modeled by Eq.(2.1) with internal noise, Eq.(3.1), the normal FR Eq.(1.1) holds. For this purpose we first demonstrate that these systems obey what we call the *Fluctuation-Dissipation Relation of the first kind* (FDR I)

$$\langle x(t) \rangle = \frac{F}{2k_B T} \langle x^2(t) \rangle_0 \;, \tag{3.2}$$

where the subscript "0" implies that the MSD $\langle x^2(t) \rangle_0$ is evaluated in the absence of the force $F$. Note that this relation is slightly different from what is defined in [34] as FDR I, but Eq.(3.2) appears to be suitably adapted to characterize anomalous dynamics [30]. By choosing $x(0)=0$, we obtain for the Laplace transform $\tilde{x}(s) = \int_0^\infty x(t)e^{-st}dt$ of Eq.(2.1) the expression

$$\tilde{x}(s) = \frac{F}{m} \frac{1}{s^2 \tilde{\gamma}(s)} + \frac{\tilde{\xi}(s)}{s\tilde{\gamma}(s)} \;. \tag{3.3}$$

After performing an inverse Laplace transformation of Eq.(3.3) we get

$$x(t) = \langle x(t) \rangle + \int_0^t dt'\, \xi(t') H(t-t') \;, \qquad \langle x(t) \rangle = \frac{F}{m}\int_0^t d\tau\, H(\tau) \;, \tag{3.4}$$

where $H(t)$ is the inverse Laplace transform of $\tilde{H}(s) = (s\tilde{\gamma}(s))^{-1}$. From Eqs.(3.4) we get

$$\sigma_x^2(t) = \langle x^2(t)\rangle_0 = 2\frac{k_BT}{m}\int_0^t d\tau_1 H(\tau_1)M(\tau_1) \quad , \tag{3.5}$$

where $M(t) = \int_0^t dt' H(t')\gamma(t-t')$. Its Laplace transform is $\tilde{M}(s) = \tilde{H}(s)\tilde{\gamma}(s) = 1/s$, therefore, $M(t) = 1$. Equation (3.5) then gives

$$\sigma_x^2(t) = \langle x^2(t)\rangle_0 = 2\frac{k_BT}{m}\int_0^t d\tau H(\tau) \quad . \tag{3.6}$$

By comparing the expressions for MD, Eq.(3.4), and MSD, Eq.(3.6), we arrive at FDR I, Eq.(3.2). In particular, after plugging Eqs.(3.4) and (3.6) into Eq.(2.4) we obtain the normal FR Eq.(1.1). Thus, for systems driven by internal Gaussian noise and described by Eq.(2.1) the existence of FDR II leads to the existence of FDR I and, as a consequence, the normal work FR holds. For the underdamped random motion of a harmonic oscillator subjected to Gaussian noise the validity of the normal FR was shown already in [35], [36]. In a more general context the validity of FRs for non-Markovian dynamics has been proved in [37], [38]. However, for external noise the situation is very different, as we will explore below.

**4. External noise case: MSD and fluctuation relations**

For external noise the FDR II, Eq.(3.1), does not hold. We consider a generic case of a system driven by external noise, namely, we assume that there is no time delay in the friction term, $\gamma(t-t') = 2\gamma\delta(t-t')$, however, $\xi(t)$ is not necessarily a white (delta-correlated) noise. Then Eq.(2.1) takes the form

$$\frac{dx}{dt} = \frac{F}{m\gamma} + \frac{\xi(t)}{\gamma} \quad . \tag{4.1}$$

Solving Eq.(4.1) with the initial condition $x(0) = 0$ gives

$$x(t) = \langle x(t)\rangle + \gamma^{-1}\int_0^t dt' \xi(t') \quad , \tag{4.2}$$

where the MD and the MSD read

$$\langle x(t)\rangle = \frac{Ft}{m\gamma} \quad , \tag{4.3}$$

$$\sigma_x^2(t) = 2\gamma^{-2}\int_0^t d\tau(t-\tau)g(\tau) \quad . \tag{4.4}$$

To analyze the FRs, we now need to specify the ACF $g(\tau)$ of the random noise. It is a well-known fact that the diffusion anomalies in a Gaussian system's dynamics come from the long power-law tails of the ACF [39]. As a consequence, in that case we expect anomalous forms of FRs. In our calculations we use the model shape of the ACF

$$g(|\tau|) = \begin{cases} g_1(|\tau|), & |\tau| \leq \Delta \\ g_2(|\tau|), & |\tau| > \Delta \end{cases} \quad , \tag{4.5}$$

where $g_1(0) > 0$, $g_1(\Delta) = g_2(\Delta)$, and

$$g_2(\tau) = C_\beta\left(\frac{\Delta}{\tau}\right)^\beta \quad , \tag{4.6}$$

where $0 < \beta < \infty$, $C_\beta > 0$ for persistent noise, whereas $1 < \beta < \infty$, $C_\beta < 0$ for antipersistent noise. Note that the condition $\int_0^\infty g(\tau)d\tau \geq 0$ must be fulfilled, otherwise the fundamental property of the non-negativity of the spectral density, $\tilde{g}(\omega) = \int_{-\infty}^\infty dt g(t)e^{i\omega t} = 2\int_0^\infty dt g(t)\cos\omega t \geq 0$, is violated [40]; this property, in particular, explains why $\beta$ must be larger than 1 for the antipersistent case. The explicit form of $g_1(\tau)$ becomes unimportant if we are interested in the long-time behaviour of the MSD at $t \gg \Delta$. Moreover, the exact form of $g_2(\tau)$ at $\tau \approx \Delta$ is also unimportant in that case. The MSD is obtained after plugging Eqs.(4.5) and (4.6) into Eq.(4.4). Then, using Eqs.(4.3) and (2.4) we get the work FR. The calculations are straightforward, and our results are summarized in Table I up to prefactors.

*4.1. Persistent external noise*
*4.1.1. $0 < \beta < 1$.* In this case the system under consideration behaves superdiffusively, and a respective explicit time dependence shows up in the work FR, which drastically changes the symmetry properties of the work PDFs: By increasing time, the probability of a negative fluctuation of the work tends to the corresponding probability of a positive fluctuation. This behaviour is very different from that of the normal FR, and for that reason we call it an *anomalous* FR. It was shown recently that asymptotically large positive and negative fluctuations of work are equally probable for Lévy flights [22], [24], which is a superdiffusive process caused by external noise possessing an alpha-stable Lévy probability law. Note that for Lévy flights the MSD does not exist; thus to characterize diffusion properties one has to use another characteristics such as, e.g., fractional moments of the order less than the Lévy index [41]. These two examples, namely Lévy flights and Gaussian systems subjected to persistent external noise with very slowly decaying ACF, allow us to make a guess that (asymptotically) equal positive and negative fluctuations of work may be observed in systems displaying superdiffusive behaviour. However, such behaviour might be a necessary but cannot be a sufficient condition. Indeed, for example, the system with antipersistent internal noise exhibits superdiffusion as well; at the same time, as we have shown in Section 3, the normal FR holds.

*4.1.2. $1 < \beta < \infty$.* This is a normal diffusion regime. Correspondingly, the symmetry property of the work PDF remains the same as for the normal FR. By introducing an effective temperature, $k_B T_{eff} = mD/\gamma$, the normal FR retains its form with $T$ substituted by $T_{eff}$. This is a *generalized* form of the FR as referred to in the introduction.

*4.1.3. $\beta = 1$.* Similar to the case $0 < \beta < 1$, the probability of a negative fluctuation tends to the probability of a positive fluctuation in the course of time. Interestingly, this tendency is logarithmically slow. This "weakly" anomalous FR represents an intermediate case between the anomalous FR and the generalized FR.

Thus, while the asymptotics of the external noise ACF becomes steeper (that is, with $\beta$ increasing) the MSD exhibits a characteristic transition from superdiffusive behaviour at $\beta < 1$ to normal diffusive behaviour at $\beta > 1$ via an intermediate $t \log t$ – behaviour at $\beta = 1$. Correspondingly, the work FR is anomalous for $0 < \beta \leq 1$ and "almost" normal (generalized) for $\beta > 1$. Similar transition scenarios for the MSD have been found for the generalized Langevin dynamics studied in [39], as well as for continuous time random walk models [42], [43].

*4.2. Antipersistent external noise*
First of all, we note that if $\int_0^\infty g(\tau)d\tau > 0$, the MSD yields normal diffusion in the long-time limit, and, similar to the persistent case with $1 < \beta < \infty$, we get a generalized FR. The situation changes in the "pure" antipersistent case, i.e., $\int_0^\infty g(\tau)d\tau = 0$. This condition implies that the spectrum of the noise, i.e., the Fourier transform $\tilde{g}(\omega)$ of the ACF has its minimum at the origin, $\tilde{g}(0) = 0$. A popular example is fractional Gaussian noise that is the (formal) derivative of antipersistent fractional Brownian motion [44].

*4.2.1. $1 < \beta < 2$.* The system behaves subdiffusively leading to an anomalous work FR: with increasing time the ratio of probabilities for positive and negative work grows to infinity, thus negative fluctuations die out in the course of time. This situation is opposite to the superdiffusive behaviour in the persistent case for $0 < \beta < 1$. However, in analogy to this superdiffusive case we may also speculate that subdiffusive behaviour is a necessary condition to observe such a suppression of negative fluctuations but not a sufficient one. Indeed, subdiffusive behaviour is inherent to systems subjected to internal persistent noise; however, in this case the normal FR holds [24].

*4.2.2. $2 < \beta < \infty$.* The MSD tends to a constant value, and the process $x(t)$ (and therefore $W(t)$) approaches its stationary counterpart. This is similar to a localization phenomenon as it has been studied by Golosov for random walks in random environments [45]. In the work FR negative fluctuations die out in the course of time even faster than in the previous subdiffusive case.

*4.2.3. $\beta = 2$.* For the MSD we have a behaviour that is intermediate between subdiffusion and approaching a constant in the long-time limit, which is logarithmic diffusion. Similar transition scenarios have again been reported in [39]. The work FR behaves accordingly: negative fluctuations die out faster than in the subdiffusive case but slower than in the stationary state.

## 5. Logarithmic factors at intermediate times

One can see that at the two values of the exponent $\beta$ characterizing long-time decay of the noise ACF, namely, $\beta = 1$ in the persistent case and $\beta = 2$ in the antipersistent case, specific logarithmic terms (log-factors) appear in the MSD and in the work FR in the long-time limit. In the persistent case $\beta = 1$ is the transition point between superdiffusion ($\beta < 1$) and normal diffusion ($\beta > 1$). In the antipersistent case $\beta = 2$ is the transition point between subdiffusion ($\beta < 2$) and stationary behaviour ($\beta > 2$). In this section we show that these log-factors actually appear not only at the two points 1 and 2 on the $\beta$-axis but also in the the neighbourhoods of $\beta$ around 1 and 2, although only for *intermediate* times. Besides clarifying the mechanism of the appearance of these log-factors, this finding indicates that they might be detectable in experiments where the value of $\beta$ is not exactly 1 or 2, but close enough to these points. Analogous survival of logarithmic terms has been found for subdiffusion in weakly chaotic maps, based on a continuous time random walk analysis [46], [47].

*5.1. Persistent external noise*

We plug Eqs.(4.5) and (4.6) into Eq.(4.4) and retain the main contributions to the MSD at $t > \Delta$ and $\beta \approx 1$:

$$\sigma_x^2(t) \approx 2\frac{C_\beta \Delta^2}{\gamma^2}\left[\frac{1}{(1-\beta)(2-\beta)}\left(\frac{t}{\Delta}\right)^{2-\beta} - \frac{t}{(1-\beta)\Delta}\right] . \qquad (5.1)$$

Assuming first that $\beta$ is less than 1, we introduce $\varepsilon = 1 - \beta$, $\varepsilon \ll 1$, and transform Eq.(5.1) into

$$\sigma_x^2(t) \approx \frac{2C_1\Delta^2}{\varepsilon\gamma^2}\left[\frac{t}{\Delta}e^{\varepsilon\ln(t/\Delta)} - \frac{t}{\Delta}\right] . \qquad (5.2)$$

Now, one can see that within the time interval

$$0 \leq \ln\left(\frac{t}{\Delta}\right) \ll \frac{1}{\varepsilon} , \qquad (5.3)$$

we can expand the exponential function in (5.2), thus getting

$$\sigma_x^2(t) \approx \frac{2C_1\Delta}{\gamma^2}t\ln\left(\frac{t}{\Delta}\right) \qquad (5.3)$$

within the time interval $\Delta < t < \exp(1/(1-\beta))$. Equation (5.3) gives exactly the MSD obtained with Eqs.(4.4)-(4.6) at $\beta = 1$. From Eq.(5.3) one can see that if $\beta$ approaches $\beta = 1$ from below, the MSD yields $t \cdot \ln t$ – behaviour at intermediate times. At longer times the power law regime $\sigma_x^2(t) \propto t^{2-\beta}$ survives. With $\beta$ getting closer and closer to 1, the time domain exhibiting this logarithmic term expands, and at $\beta = 1$ it spans to infinity. The analysis for $\beta$ above 1 is completely identical with $\varepsilon = \beta - 1$, $\varepsilon \ll 1$, $\Delta < t < \exp(1/(\beta - 1))$. We have thus shown the existence of the $t \cdot \ln t$ – behaviour of the MSD within the time interval $\Delta < t < t_{c1}$, $t_{c1} = \exp(1/|1-\beta|)$. Accordingly, within this interval the work FR yields $\ln(p(W,t)/p(-W,t)) \sim W/\ln(t/\Delta)$, as for $\beta = 1$.

*5.2. Antipersistent external noise*

We consider the behaviour of the MSD in the vicinity of $\beta = 2$. In complete analogy to the persistent case we get

$$\sigma_x^2(t) \approx \frac{2|C_2|\Delta^2}{\gamma^2}\ln\left(\frac{t}{\Delta}\right) \quad (5.6)$$

for intermediate times

$$\Delta < t < t_{c2} = \Delta\exp\left(\frac{1}{|\beta-2|}\right). \quad (5.7)$$

Similarly to the persistent noise case we conclude that if the exponent $\beta$ is close to 2, then at intermediate times $\Delta < t < t_{c2} = \Delta\exp(1/|\beta-2|)$ the MSD grows logarithmically with time, as for $\beta$ equal to 2. For $t > t_{c2}$ the MSD grows like $t^{2-\beta}$ at $1 < \beta < 2$ and tends to a constant at $\beta > 2$. As $\beta$ approaches 2, the value of $t_{c2}$ increases, and at $\beta = 2$ the time domain in which the MSD exhibits logarithmic behaviour expands to infinite times. Accordingly, for intermediate times the work FR yields $\ln(p(W,t)/p(-W,t)) \sim Wt/\ln(t/\Delta)$, as for $\beta = 2$.

## 6. Conclusions

Our analysis sheds light on the interplay between the existence of normal FRs, Eq.(1.1), and the existence of FDRs for the paradigmatic example of Gaussian stochastic dynamics with correlated noise. In case of internal noise, which means that FDR II holds, we have found that this implies the validity of FDR I and hence the existence of normal transient work FRs. For external noise where FDR II is broken, however, the situation is more complicated: Here we have found both the existence of normal and what we call anomalous FRs depending on whether the noise is persistently or antipersistently correlated, and in addition depending on control parameters, cf. Table I. The whole situation is summarized in Fig.1, which highlights an interesting connection between FDR I and II, and the precise form of FRs in Gaussian stochastic dynamics. Note that, according to this figure, an experimental test yielding an anomalous transient work FR for this type of systems would immediately rule out the existence of FDR I and II.

We remark at this point that there is no unique denotation of different forms of FRs in the literature. In our case, we called any transient work FR that is different from the strict functional form of the "normal" FR Eq.(1.1) an "anomalous" FR. This generalizes the classification suggested in [48], where an anomalous fluctuation property referred to power law PDFs only. That way, the "extended" and "generalized" FRs of [19], [20], [21], [48] all form sub-classes of what we call anomalous FRs. A more detailed classification scheme would be desirable.

Specifically, we have reported the existence of several new forms of anomalous FRs, cf. Table I: For two of them the large deviation form Eq.(1.1) is still preserved, however, by featuring time-dependent pre-factors on the right hand side. While the case of decreasing slope was already reported in [24], the analytical case with increasing slope is new. Also new are the two cases with multiplicative logarithmic corrections, as well as the one case exhibiting localization in the MSD leading to a linear time dependence in the FR. We have also shown that the logarithmic factors persist for long times around the transition points between different types of diffusive behaviour, hence their existence is non-negligible and should be accounted for in investigations of anomalous dynamics. We emphasize that these anomalous FRs do not constitute a violation of the Second Law of Thermodynamics: firstly, the average value of work (which in this case equals entropy production) is still positive; and secondly, we only observe such anomalies in case of systems driven externally.

In real experiments as well as in computer simulations these forms of anomalous FRs could be detected by plotting the fluctuation ratio, i.e., the logarithm of the PDF ratio on the left hand side of Eq.(1.1), as a function of the work at different instances of time. Such an analysis has already been performed for Monte Carlo simulations of a lattice-gas analogue of a non-Newtonian glassy fluid driven by a constant, uniform force field [23]. In this paper a fluctuation ratio with a slope that increased with increasing time has been reported with a subdiffusive regime for intermediate times and weak fields, which is in qualitative agreement with our "pure" antipersistent regime for a certain range of parameters. Even more, for strong fields it was found numerically that the slope of the fluctuation ratio decreased with increasing time while the system exhibited superdiffusive behaviour at intermediate times, which again is in qualitative agreement with our persistent case for a certain range of parameters. An existence of logarithmic corrections in the MSD has also been suggested by respective data analysis. To explore whether a detailed physical relation between the model studied in this paper and the generalized Langevin dynamics considered in our work can be established, remains an interesting open question. Hints on the existence of an anomalous FR have also been obtained for the anomalous dynamics of biological cells migrating under chemical gradients [27]. Note that such cells consist of a complicated biopolymer gel exhibiting glassy rheological properties. We finally highlight again the generalization of normal FRs in [19], [20] for glassy systems. Altogether one may thus conjecture that glassy dynamics might provide an ideal testing ground in order to look for the anomalous FRs suggested by our theoretical analysis in experiments.

**Acknowledgements**

We thank the London Mathematical Society for financial support in form of a travel grant for AVC under reference No.51005.

**Table I. Different diffusion and work fluctuation relation regimes at $t \gg \Delta$.**

| $\beta$ | Persistent | | Antipersistent, $\int_0^\infty d\tau g(\tau) = 0$ | |
|---|---|---|---|---|
| | MSD | $\ln(p(W,t)/p(-W,t))$ | MSD | $\ln(p(W,t)/p(-W,t))$ |
| $0 < \beta < 1$ | $\propto t^{2-\beta}$ | $\propto \dfrac{W}{t^{1-\beta}}$ | The regime does not exist | |
| $\beta = 1$ | $\propto t\ln\left(\dfrac{t}{\Delta}\right)$ | $\propto \dfrac{W}{\ln(t/\Delta)}$ | | |
| $1 < \beta < 2$ | $\sim 2Dt$ $D = \dfrac{1}{\gamma^2}\int_0^\infty d\tau\, g(\tau)$ | $\sim \dfrac{W}{k_B T_{eff}}$ $k_B T_{eff} = mD/\gamma$ | $\propto t^{2-\beta}$ | $\propto Wt^{\beta-1}$ |
| $\beta = 2$ | | | $\propto \ln\left(\dfrac{t}{\Delta}\right)$ | $\propto \dfrac{Wt}{\ln(t/\Delta)}$ |
| $\beta > 2$ | | | $\sim \dfrac{2}{\gamma^2}\left|\int_0^\infty d\tau\tau g(\tau)\right|$ | $\sim \dfrac{\gamma}{m\left|\int_0^\infty d\tau\tau g(\tau)\right|}Wt$ |

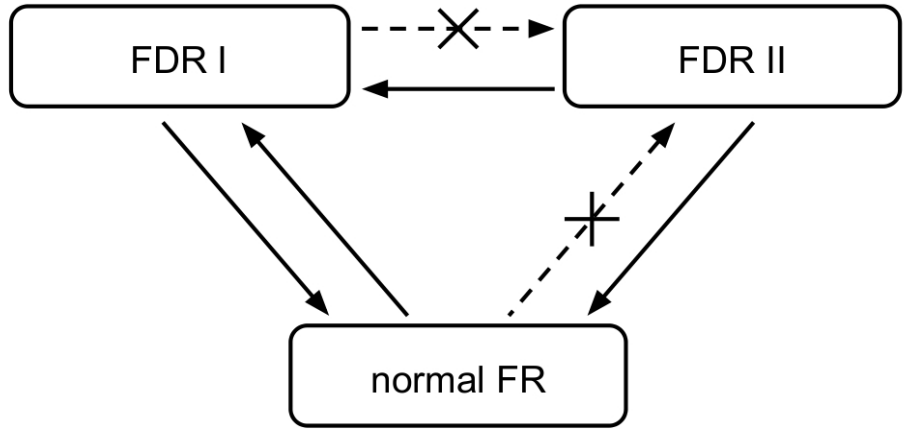

**Fig.1.** Logical relation between Fluctuation-Dissipation Relations and the normal Fluctuation Relation Eq.(1.1) for Gaussian stochastic systems with constant drift.